\documentclass[conference]{IEEEtran}
\IEEEoverridecommandlockouts
\usepackage{cite}
\usepackage{amsmath,amssymb,amsfonts}
\usepackage{algorithmic}
\usepackage{graphicx}
\usepackage{textcomp}
\usepackage{xcolor}
\usepackage{multirow}
\usepackage{booktabs} 

\usepackage{subcaption} 
\usepackage{mwe} 

\usepackage{tikz}
\usepackage{textcomp}
\usepackage[hidelinks]{hyperref}

\newcommand\copyrighttext{%
  \footnotesize \textcopyright 2023 IEEE. Personal use of this material is permitted.
  Permission from IEEE must be obtained for all other uses, in any current or future
  media, including reprinting/republishing this material for advertising or promotional
  purposes, creating new collective works, for resale or redistribution to servers or
  lists, or reuse of any copyrighted component of this work in other works.
  DOI: \href{https://doi.org/10.1109/ICOIN56518.2023.10048928}{10.1109/ICOIN56518.2023.10048928}}
\newcommand\copyrightnotice{%
\begin{tikzpicture}[remember picture,overlay]
\node[anchor=south,yshift=10] at (current page.south) {\fbox{\parbox{\dimexpr\textwidth-\fboxsep-\fboxrule\relax}{\copyrighttext}}};
\end{tikzpicture}%
}

\def\BibTeX{{\rm B\kern-.05em{\sc i\kern-.025em b}\kern-.08em
    T\kern-.1667em\lower.7ex\hbox{E}\kern-.125emX}}
\begin{document}

\title{Transformers with Attentive Federated Aggregation for Time Series Stock Forecasting
\thanks {This work was supported by the Institute of Information and Communications Technology Planning and Evaluation (IITP) Grant funded by the Korea Government (MSIT) (Artificial Intelligence Innovation Hub) under Grant 2021-0-02068 and in part by the National Research Foundation of Korea (NRF) grant funded by the Korea government (MSIT) (No. 2020R1A4A1018607) *Dr. CS Hong is the corresponding author.
}}

\author{\IEEEauthorblockN{Chu Myaet Thwal, Ye Lin Tun, Kitae Kim, Seong-Bae Park, Choong Seon Hong\textsuperscript{*}}
\IEEEauthorblockA{\textit{Department of Computer Science and Engineering, Kyung Hee University, Yongin-si 17104, Republic of Korea} \\
Email: \{chumyaet, yelintun, glideslope, sbpark71, cshong\}@khu.ac.kr}}

\maketitle
\copyrightnotice
\begin{abstract}
Recent innovations in transformers have shown their superior performance in natural language processing (NLP) and computer vision (CV). The ability to capture long-range dependencies and interactions in sequential data has also triggered a great interest in time series modeling, leading to the widespread use of transformers in many time series applications. However, being the most common and crucial application, the adaptation of transformers to time series forecasting has remained limited, with both promising and inconsistent results. In contrast to the challenges in NLP and CV, time series problems not only add the complexity of order or temporal dependence among input sequences but also consider trend, level, and seasonality information that much of this data is valuable for decision making. The conventional training scheme has shown deficiencies regarding model overfitting, data scarcity, and privacy issues when working with transformers for a forecasting task. In this work, we propose attentive federated transformers for time series stock forecasting with better performance while preserving the privacy of participating enterprises. Empirical results on various stock data from the Yahoo! Finance website indicate the superiority of our proposed scheme in dealing with the above challenges and data heterogeneity in federated learning.
\end{abstract}


\section{Introduction}
Time series forecasting is the task of analyzing historical and current time-stamped data to make scientific predictions over a period of time that can inform future strategic decisions. Unlike other types of tasks, the future outcome of a forecasting problem is not known in advance; instead, it can only be approximated by leveraging historical data analysis. Especially when dealing with the frequently changing variables in time series data and events that cannot be controlled, it is not always possible to make an accurate prediction, and the likelihood of such forecasts might vary greatly~\cite{torres2021deep}. Hence, studies on time series forecasting have proved to be useful in a variety of contexts, including environmental, healthcare, financial, weather forecasting, and so on~\cite{lim2021time}. It is feasible to use machine learning (ML) approaches such as regression, random forest (RF), support vector machines (SVM), or artificial neural networks (ANN) and fit models on the historical data to predict future observations. However, the most crucial step in any time series forecasting problem is to develop efficient prediction models with the ability to learn from raw data to recognize the underlying hidden patterns, which the majority of ML algorithms may not be capable of by default~\cite{masini2021machine}.

\begin{figure}[t]
     \centering
     \begin{subfigure}[b]{\columnwidth}
         \centering
         \includegraphics[width=0.95\columnwidth]{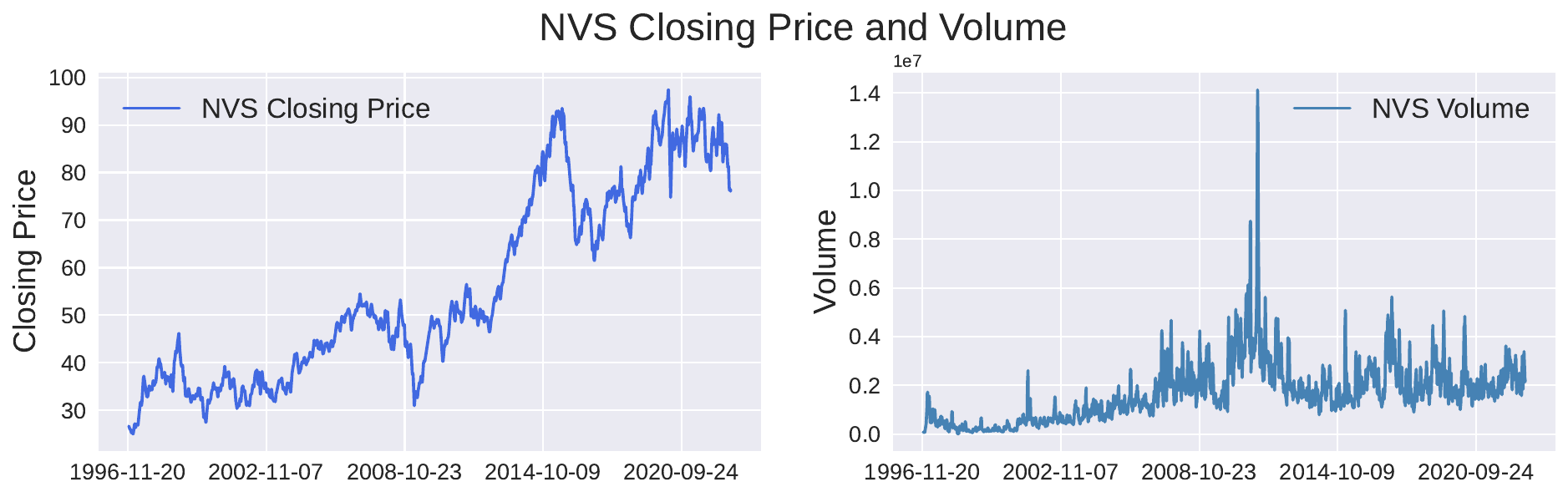}
         \label{fig:data_vis_NVS}
     \end{subfigure}
     \hfill
     \begin{subfigure}[b]{\columnwidth}
         \centering
         \includegraphics[width=0.95\columnwidth]{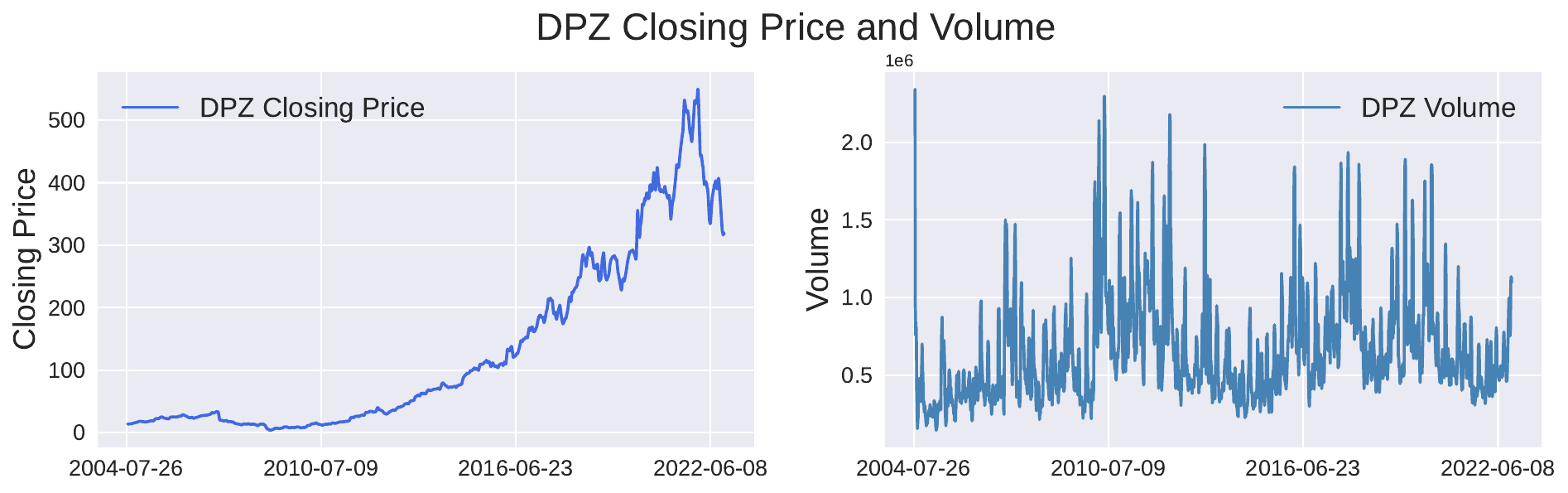}
         \label{fig:data_vis_DPZ}
     \end{subfigure}
        \caption{Visualization of daily stock trend for NVS and DPZ datasets. Values are retrieved from Yahoo! Finance website~\cite{yahoofinance}.}
        \label{fig:data_vis}
\end{figure}

\begin{figure*}[t]
	\centering
	\includegraphics[width=0.74\textwidth]{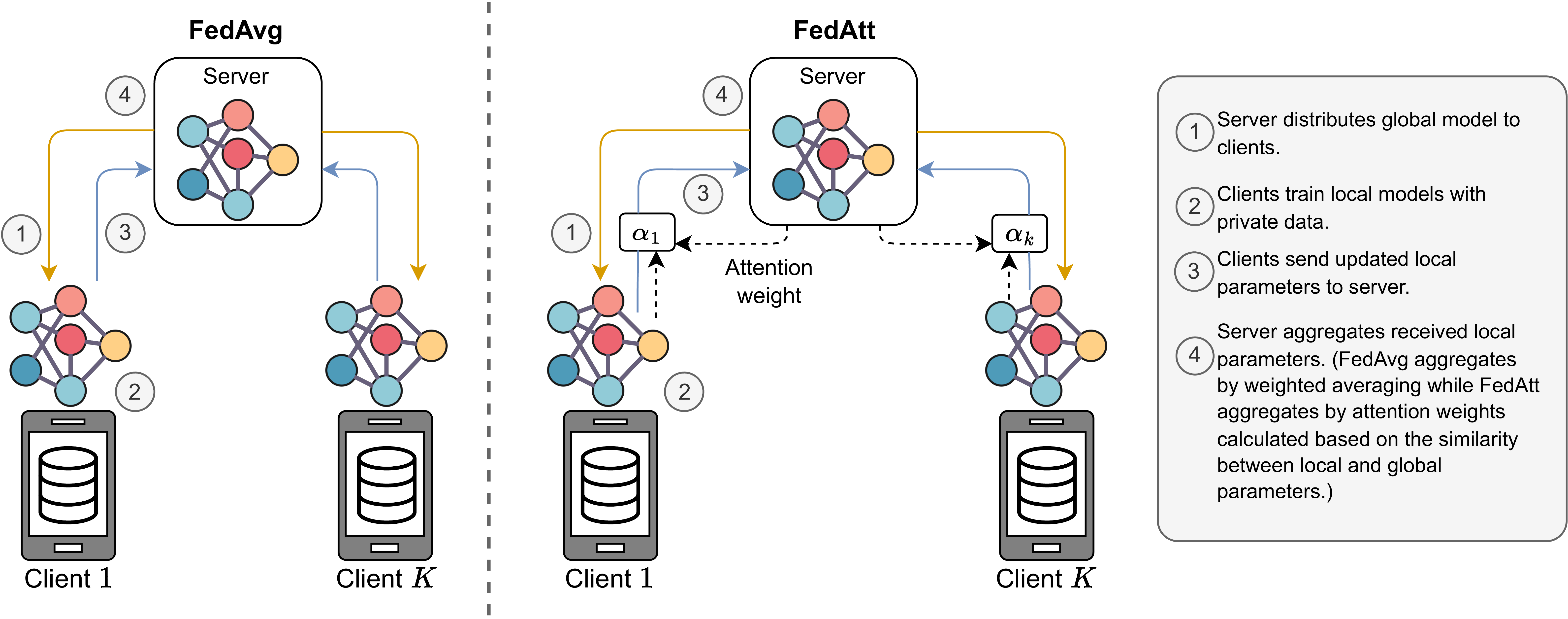}
	\caption{An overview of FedAvg and FedAtt schemes.}
	\label{fig:fed_schemes}
\end{figure*}

Recently, stock market forecasting has gained a substantial amount of attention as a result of the potential financial benefits. It is very important to yield accurate forecasting results, as stocks are the most risky and volatile investments, and crucial in the world of finance and business. Although traditional ML approaches such as extreme gradient boosting (XGBoost), autoregressive integrated moving average (ARIMA) and ANNs like long short-term memory (LSTM) have shown their potential in extrapolating stock prices, the article on developing efficient models to further minimize the forecasting error is still attracting interest from the research community. Due to the uncertainty of features involved and their complex and noisy nature, it becomes challenging to accurately forecast stock market trends. As the trend is continuously shifting under the influence of several factors, many of which remain unknown and uncontrollable, there is no consistent pattern to follow. Thus, it is difficult to apply traditional ML approaches when dealing with non-stationary stock forecasting problems. With the widespread improvement in deep learning (DL) and artificial intelligence (AI) technologies, a variety of innovative models for predicting the future trends of stocks by carefully examining the patterns of historical rates have been widely proposed and proved to be effective. Fig.~\ref{fig:data_vis} represents the visualization of stock market trends for `Novartis AG (NVS)' and `Domino's Pizza, Inc. (DPZ)' enterprises, showing the closing price and volume features. From the figure, we can observe the fluctuating nature of the market trend per enterprise.

In recent years, transformers, being the most powerful sequence modeling architectures, have gained popularity due to their outstanding performance in natural language processing (NLP) and computer vision (CV)~\cite{vaswani2017attention, devlin2018bert, dosovitskiy2020image, liu2021swin}. With the combination of positional encoding and multi-head self-attention mechanism, transformers achieve an impressive capability of parallelization and extracting semantic features from a long sequence, i.e., words or image patches. Moreover, there has been a surge of research interest in transformers for time series modeling by the fact that they can capture long-range dependencies and interactions among the sequential data. Hence, a variety of transformer-based solutions have been developed for forecasting applications as the most prevalent and crucial factor in the field of time series~\cite{wen2022transformers}. In contrast, while the positional encoding helps to maintain some ordering information, there remains a temporal information loss due to the permutation-invariant nature of the self-attention component when working with time series transformers. Consequently, when analyzing time series numerical data that lacks semantic elements, it is vital to focus on modeling the temporal changes among an ordered set of continuous points.

Nonetheless, large amount of data is required in training a data-hungry transformer with minimum overfitting issue. In some cases, it may not be possible to obtain a sufficient amount of training data when working with stock prices i.e., historical data of many enterprises is not publicly available, or only a short period of historical data is recorded, as in `Meta Platforms, Inc. (META)' and `Tripadvisor, Inc. (TRIP)' datasets, where stock values are available only from dates `2012-05-18' and `2011-12-07', respectively. In recent works, distributed machine learning techniques have proved to be effective in developing better neural networks for data-intensive applications. Federated learning (FL) is a promising approach that overcomes data scarcity and privacy challenges as it enables the collaborative learning of a shared global model by aggregating locally computed updates from distributed client models with decentralized private data~\cite{mcmahan2017communication}. To the best of our knowledge, the adaptation of federated transformers to time series problems has remained limited, and thus, our motivation is \textbf{to explore whether federated transformers are effective for time series forecasting tasks}.

In this work, we develop a transformer-based architecture for time series forecasting, especially addressing challenges in the time series stock market. We preserve the temporal information of time series data by embedding the vector representation for time into the input sequence. Specifically, we train our model on historical daily stock data where the input integer (day) is used as the time feature for Time2Vec representation~\cite{kazemi2019time2vec} and a number of transformer encoders are stacked above to predict the output trend. We explore the effectiveness of our time series forecasting transformer in FL scenarios to enhance accuracy while coping with data heterogeneity, scarcity, and privacy issues. We combine our model with attentive federated learning (FedAtt)~\cite{ji2019learning} and analyze the efficacy of our proposed scheme in comparison with decentralized local training (SOLO) and federated averaging (FedAvg) baselines.
Main contributions of this work are:
\begin{itemize}
\item We develop a time series transformer based on the multi-head self-attention mechanism to effectively forecast the future trends of closing price on the global stock market.
\item We utilize the federated learning scheme, specifically the attentive aggregation mechanism (FedAtt), to enable the collaborative learning of our models by leveraging the distributed historical data of different enterprises.
\item We analyze the performance of our proposed scheme by conducting an evaluation on the public stock data retrieved from Yahoo! Finance website~\cite{yahoofinance}, in comparison with two baselines, i.e., the decentralized training of local models (SOLO) and the federated averaging (FedAvg).
\end{itemize}

The rest of this paper is organized as follows: in section~\ref{related_work}, we provide a brief literature review on existing time series forecasting models and an overview of the FL paradigm. We present our proposed scheme in section~\ref{system_model}, by explaining the main components of the time series transformer and how we integrate it into the FL paradigm. Experimental details and simulation results are discussed in section~\ref{experiments}. Finally, we conclude our work in section~\ref{conclusion}.


\section{Related Work} \label{related_work}


\subsection{Time Series Forecasting}

\begin{figure}[t]
	\centering
	\includegraphics[width=\columnwidth]{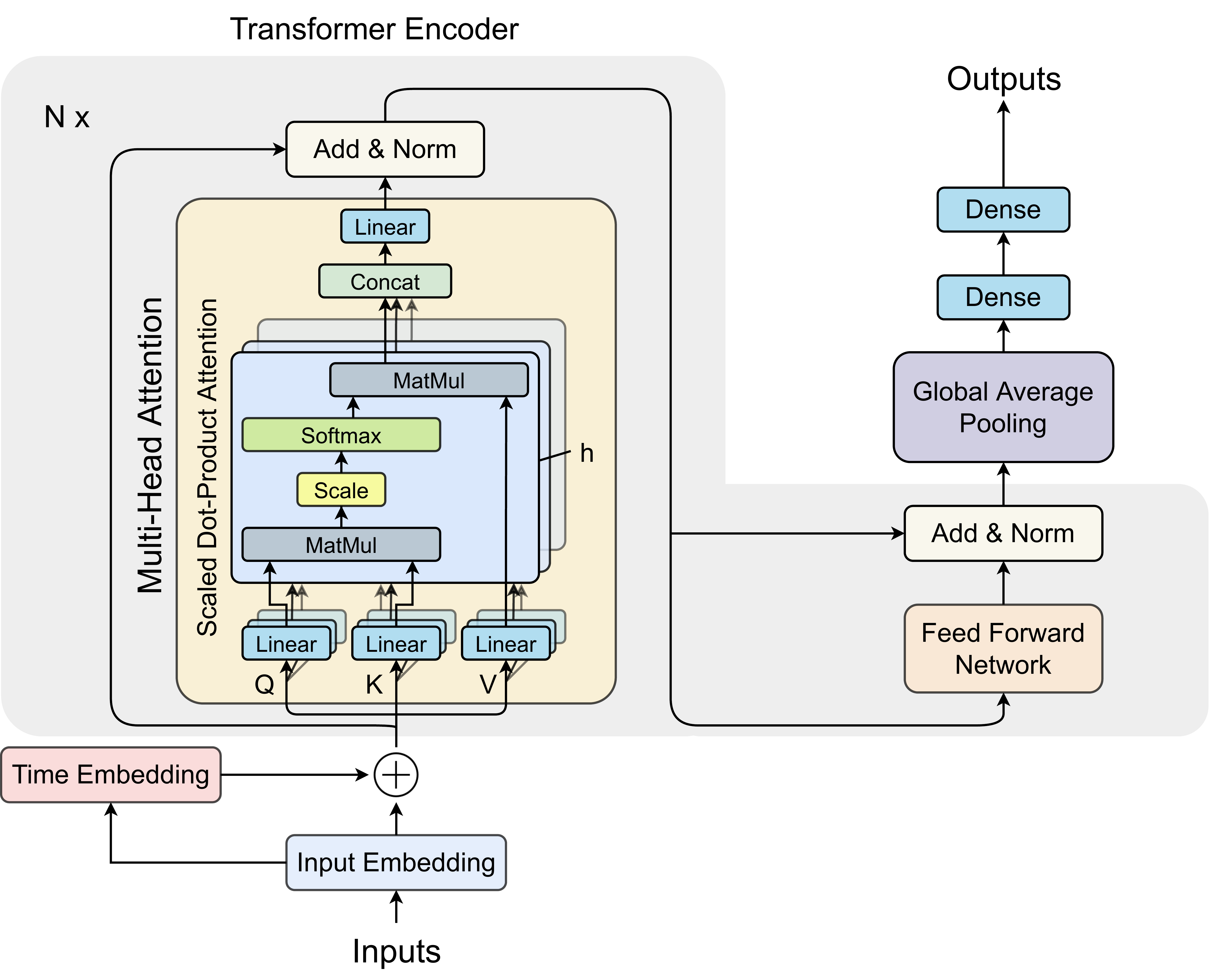}
	\caption{Architecture of a time series forecasting transformer.}
	\label{fig:system_model}
\end{figure}

Most common and popular baselines for time series forecasting problems are the ARIMA models based on which several works have been widely proposed in stock forecasting~\cite{areekul2009notice, banerjee2014forecasting}. As combining econometric models like ARIMA with ANNs yields encouraging results in developing forecasting models for financial markets, LSTM, which is an advanced type of neural network specifically developed for processing sequential data, has gained popolarity~\cite{roondiwala2017predicting, baek2018modaugnet, tan2019ensemble, moghar2020stock, feng2018enhancing, wang2021clvsa}. In~\cite{patel2015predicting}, the authors used a wide range of ML models such as Naive Bayes classifiers, RF, SVM, and ANN to forecast the trends of the Indian stock market. Their empirical results show the superiority of RF compared to the other models. The authors in \cite{pyo2017predictability} applied SVM and ANN to forecast the direction of the Korean stock price. The work in \cite{borovkova2019ensemble} proposed an ensemble of LSTM models to estimate the future trends of large-cap US stocks. Inputs to the model consist of price-based features in combination with other advanced technical indicators, and it is shown that their proposed approach outperforms the regression models. To denoise historical stock data, extract its features, and create a stock forecasting model, \cite{qiu2020forecasting} suggested a wavelet transform based on LSTM and an attention mechanism.

In \cite{hoseinzade2019cnnpred, tsantekidis2017forecasting, selvin2017stock}, the authors exploited the convolutional neural network (CNN) and recurrent neural network (RNN) architectures for stock price forecasting tasks. Reinforcement learning (RL) has also been used to improve the performance of stock forecasting as proposed in~\cite{wang2019alphastock}. Aside from image generation, generative adversarial networks (GANs) \cite{goodfellow2020generative} have proved to be useful for time series forecasting. \cite{zhou2018stock} proposed high-frequency stock market forecasting by building a GAN model where LSTM and CNN models are used as the generator and the discriminator. Multi-layer perceptron (MLP) has also been employed as the discriminator in GAN to forecast for predicting stock values~\cite{zhang2019stock}. Moreover, the most recent and efficient time series forecasting transformers include Informer~\cite{zhou2021informer}, Autoformer~\cite{wu2021autoformer}, FEDformer~\cite{zhou2022fedformer}, ETSformer~\cite{woo2022etsformer}, Pyraformer~\cite{liu2021pyraformer} and so on. Although transformers demonstrate promising results in time series forecasting, further studies are required to tackle the challenges of financial time series forecasting tasks.


\subsection{Federated Learning}


Federated learning (FL) enables the collaborative training of a machine learning algorithm on private data distributed across multiple decentralized devices. The earliest and most common FL framework is federated averaging (FedAvg)~\cite{mcmahan2017communication}, which is an iterative model averaging process that contains four key steps in each iteration, as shown in the left side of Fig.~\ref{fig:fed_schemes}. First, the server randomly initializes a global model and distributes it to the participating clients. Each client updates its local model by using stochastic gradient descent (SGD) on private training data. After that, the updated local models are sent to the server for aggregation. Finally, the server performs weighted averaging on the received local model parameters to update the global model for the next iteration. These steps are repeated until the global model converges.

\section{System Architecture} \label{system_model}

\subsection{Time2Vector Embeddings}
When processing sequential or time series data with a transformer, it is hard to extract the sequential and temporal dependencies as the input sequences are sent through the encoder at once. In NLP transformers~\cite{vaswani2017attention, devlin2018bert}, positional encoding is employed prior to the transformer encoder to embed word order in the input sequence, providing positional information to the model. Similarly, in order to implement a time series stock forecasting transformer, it is crucial to encode the time feature hidden in the stock data and incorporate it with the other input features, i.e., four price features in our task (Open, High, Low, and Close). Without time embeddings, a transformer model would not be able to obtain any information regarding the temporal order of stock values.

Time2Vec~\cite{kazemi2019time2vec} is a model-agnostic vector representation, used to encode temporal features in the form of vector representations. Authors in~\cite{kazemi2019time2vec} observed that both periodic and non-periodic patterns are important for a meaningful temporal representation, as well as a time representation should be invariant to time rescaling, i.e., it has to retain integrity with different time increments. By combining the concepts of periodic and non-periodic patterns with the idea of invariance to temporal rescaling, we initialize Time2Vec layer as a time embedding layer in our model prior to the transformer encoder:

\begin{equation}
    \textbf{t2v} (\tau)[i] =
    \begin{cases}
    \omega_i\tau+\phi_i,& \text{if } i=0,\\
    \mathcal{F}(\omega_i\tau+\phi_i),& \text{if } 1 \leq i \leq k,
    \end{cases}
\end{equation}
where \textbf{t2v} denotes the time to vector representation with two components: $\omega_i\tau+\phi_i$ for linear or non-periodic feature and $\mathcal{F}(\omega_i\tau+\phi_i)$ for periodic feature of the time vector. Thus, the time embedding of an input sequence is obtained and forwarded to the transformer encoder. 

\subsection{Transformer Encoder}
The input embedding, incorporated with the time embedding, is fed to the transformer encoder as the initial input to the self-attention (SA) layer. The SA layer separates the input into three vectors: Query $Q$, Key $K$, and Value $V$. In the case of stock data, $(Q, K, V)$ values represent the price, volume, and time features. By passing each $Q$, $K$, and $V$ through an individual linear layer, a separate linear transformation of each matrix is obtained. Thus, attention weights are calculated by taking the dot-product of $Q$ and $K$ matrices and divided by the dimension of the previous vectors, i.e., $d_k=256$ in our case, to prevent the gradient explosion. After the dot-product is calculated, $\mathrm{softmax}$ function is applied to generate a set of weights that add up to 1. In order to complete the self-attention mechanism, the transformed $V$ matrix is multiplied by the output of the $\mathrm{softmax}$ to set the attention weight at each time step. Scaled dot-product self-attention is represented by:

\begin{equation}
    SA(Q, K, V) = \mathrm{softmax}(\frac{QK^T}{\sqrt{d_k}})V.
\end{equation}

To make self-attention stronger and more efficient, multi-head self-attention (MHSA) layer is implemented by concatenating the attention weights of $h$ single SA layers. Hence, our transformer can simultaneously attend on multiple time series steps. In our architecture, we employ 12 attention heads, and the capacity to capture long-distance dependencies can be improved with the increase of attention heads $h$. Multi-head self-attention can be represented by:
\begin{equation}
    MHSA(Q, K, V) = \mathrm{Concat}(SA_1, \dots, SA_h).
\end{equation}

Each transformer encoder comprises two sub-layers: a MHSA layer and a feed-forward layer, each with a residual connection for adding the initial input. The feed-forward layer consists of two dense layers with a ReLU activation in between. The output of each sub-layer is normalized to stabilize and speed up the training. $N$ layers of transformer encoders are stacked before the global average pooling layer and the final regression layers. Fig.~\ref{fig:system_model} shows an overview of our time series forecasting transformer architecture. Different from the NLP transformers for 2-dimensional input sequences, our time series transformer can handle a 3-dimensional time series sequence with the help of the time embedding layer.

\subsection{Attentive Federated Learning}
To facilitate the performance of our time series transformer in low-data regimes, we integrate our model with an attentive federated learning (FedAtt) scheme~\cite{ji2019learning}. In addition to the vanilla FedAvg scheme, which simply averages the local model updates and ignores the importance of each client, FedAtt introduces an attention mechanism in the process of model aggregation. To find the optimized global model, FedAtt measures the importance of each participating client by calculating the similarity between the parameters of the global model and the corresponding local model, i.e., attention weights. The right side of Fig.~\ref{fig:fed_schemes} depicts the training overview of the FedAtt scheme. By combing our time series transformer with the FedAtt scheme, it optimizes the distance between the global model and client models in parameter space to learn a well-generalized global model for different enterprises.

\section{Experiments} \label{experiments}

\begin{figure}[t]
     \centering
     \begin{subfigure}[b]{\columnwidth}
         \centering
         \includegraphics[width=0.98\columnwidth]{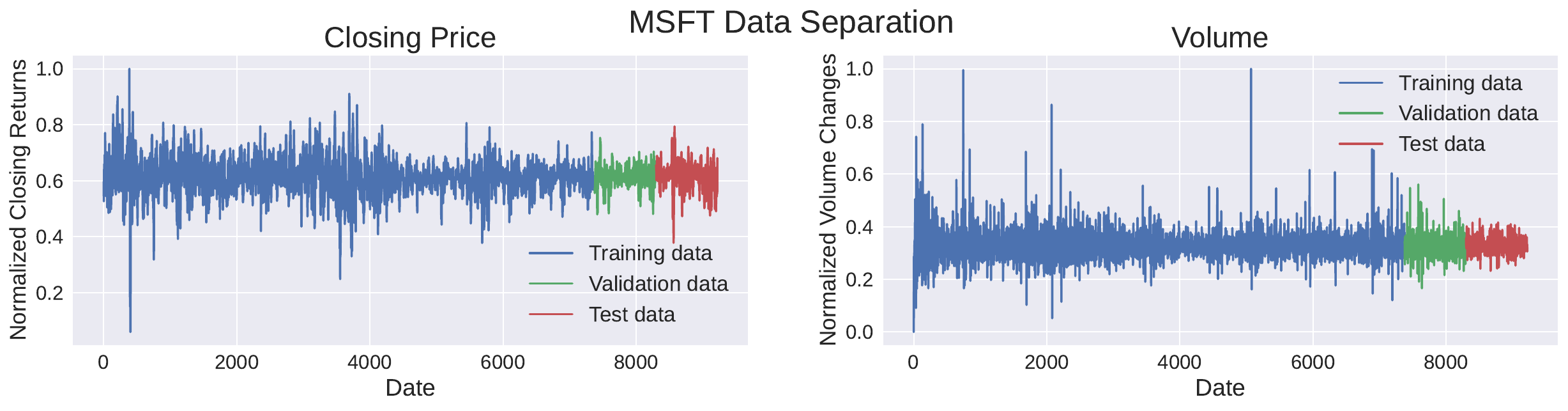}
         \label{fig:data_sepMSFT}
     \end{subfigure}
     \hfill
     \begin{subfigure}[b]{\columnwidth}
         \centering
         \includegraphics[width=0.98\columnwidth]{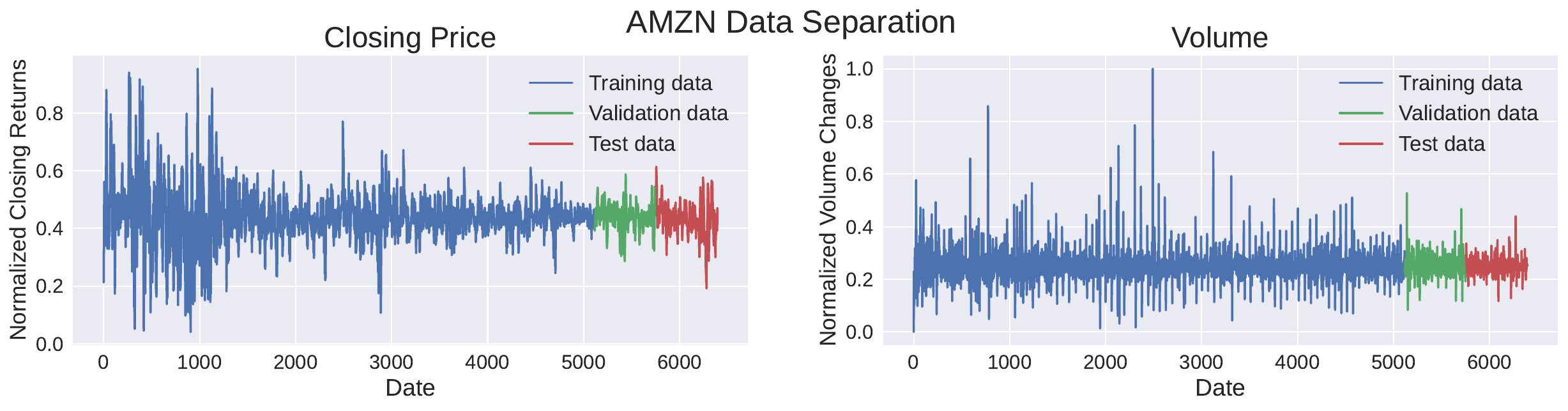}
         \label{fig:data_sepAMZN}
     \end{subfigure}
        \caption{Visualizing data separation of MSFT and AMZN datasets for time series forecasting. Best viewed in color.}
        \label{fig:data_sep}
\end{figure}

\begin{table}[b]
\centering
\caption{Performance comparison on the test data. The best values are marked in bold.}
\label{metrics}
\resizebox{\columnwidth}{!}{%
\begin{tabular}{c|c|ccc} 
\toprule
\multirow{2}{*}{Dataset}                                                                                & Model: Time Series Transformer & \multicolumn{3}{c}{Evaluation Metrics}  \\ 
\cline{2-5}
                                                                                                        & Method                         & MSE    & MAE    & MAPE                  \\ 
\hline
\multirow{3}{*}{\begin{tabular}[c]{@{}c@{}}COST \\(start year: 1986)\\data points = 9,140\end{tabular}} & SOLO                           & 0.0012 & 0.0264 & 4.6558                \\
                                                                                                        & FedAvg                         & 0.0018 & 0.0314 & 5.4925                \\
                                                                                                        & FedAtt (Proposed)              & \textbf{0.0011} & \textbf{0.0231} & \textbf{3.9396}                \\ 
\hline
\multirow{3}{*}{\begin{tabular}[c]{@{}c@{}}IBM\\(start year: 1962)\\data points = 15,300\end{tabular}}  & SOLO                           & \textbf{0.0013} & \textbf{0.0251} & \textbf{4.9934}                \\
                                                                                                        & FedAvg                         & 0.0023 & 0.0371 & 7.1588                \\
                                                                                                        & FedAtt (Proposed)              & 0.0016 & 0.0300 & 6.0320                \\ 
\hline
\multirow{3}{*}{\begin{tabular}[c]{@{}c@{}}META\\(start year: 2012)\\data points = 2,617\end{tabular}}  & SOLO                           & 0.0103 & 0.0744 & 19.3494               \\
                                                                                                        & FedAvg                         & 0.0101 & 0.0728 & 19.3195               \\
                                                                                                        & FedAtt (Proposed)              & \textbf{0.0041} & \textbf{0.0497} & \textbf{11.4834}               \\ 
\hline
\multirow{3}{*}{\begin{tabular}[c]{@{}c@{}}MSFT\\(start year: 1986)\\data points = 9,221\end{tabular}}  & SOLO                           & 0.0023 & 0.0358 & 6.0712                \\
                                                                                                        & FedAvg                         & 0.0014 & 0.0297 & 5.0121                \\
                                                                                                        & FedAtt (Proposed)              & \textbf{0.0007} & \textbf{0.0200} & \textbf{3.3118}                \\ 
\hline
\multirow{3}{*}{\begin{tabular}[c]{@{}c@{}}TMUS\\(start year: 2007)\\data points = 3,899\end{tabular}}  & SOLO                           & 0.0025 & 0.0370 & 5.9582                \\
                                                                                                        & FedAvg                         & 0.0022 & 0.0392 & 6.7448                \\
                                                                                                        & FedAtt (Proposed)              & \textbf{0.0016} & \textbf{0.0341} & \textbf{5.8555}                \\
\bottomrule
\end{tabular}
}
\end{table}

In this section, we conduct extensive experiments to analyze the superior performance of our proposed time series forecasting transformer with federated attentive aggregation (FedAtt).

\subsection{Baselines}
The performance of our time series stock forecasting transformers in the FedAtt scheme is compared with two baseline approaches: SOLO, in which each client trains its local model on its own, and FedAvg, in which all clients collaboratively train a federated model by weighted averaging mechanism.

\subsection{Datasets} We retrieve the real-time historical stock price data of 45 separate global enterprises from the Yahoo! Finance website~\cite{yahoofinance}. Retrieved datasets have different starting dates, and end on the date `2022-10-25'. As we use various sizes of datasets for the data heterogeneity purpose, each dataset contains different numbers of data points. Each data point consists of 6 features, including the date of the point, the trading volume of the stock, as well as 4 price features, i.e., Open, High, Low, and Close. We apply the moving average smoothing effect with a window size of 10 days to all the features for a better forecasting result. We enhance the stationarity of our datasets by converting the volume and price features into the daily volume changes and stock returns. Thus, we can train our models with a higher validity level of forecasting results. Then, the values are min-max normalized and split into 80\% for training, 10\% for validation and 10\% for test set. Finally, the training, validation and test sets are separated into individual sequences with a length of 16 days and 5 features per sequence day i.e., Volume, Open, High, Low, and Close. Each dataset of an enterprise represents the private database of a local client with its own forecasting model. Fig~\ref{fig:data_sep} shows the visualization of data separation for `Microsoft Corporation (MSFT)' and `Amazon.com, Inc. (AMZN)' datasets. 


\begin{figure}[t]
     \centering
     \begin{subfigure}[b]{\columnwidth}
         \centering
         \includegraphics[width=0.92\columnwidth]{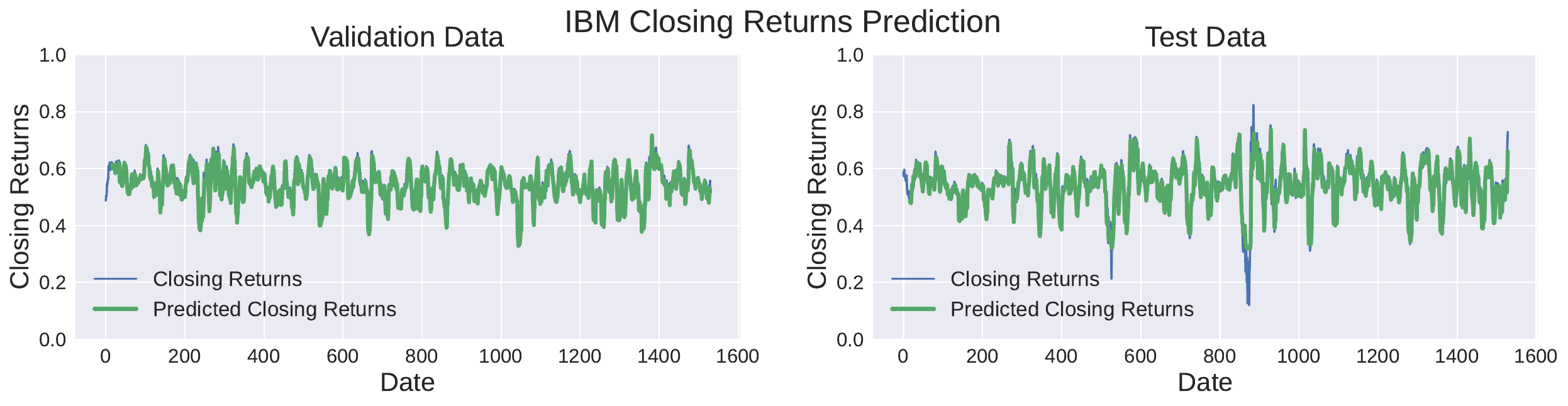}
         \caption{SOLO}
         \label{fig:fedavg_IBM}
     \end{subfigure}
     \hfill
     \begin{subfigure}[b]{\columnwidth}
         \centering
         \includegraphics[width=0.92\columnwidth]{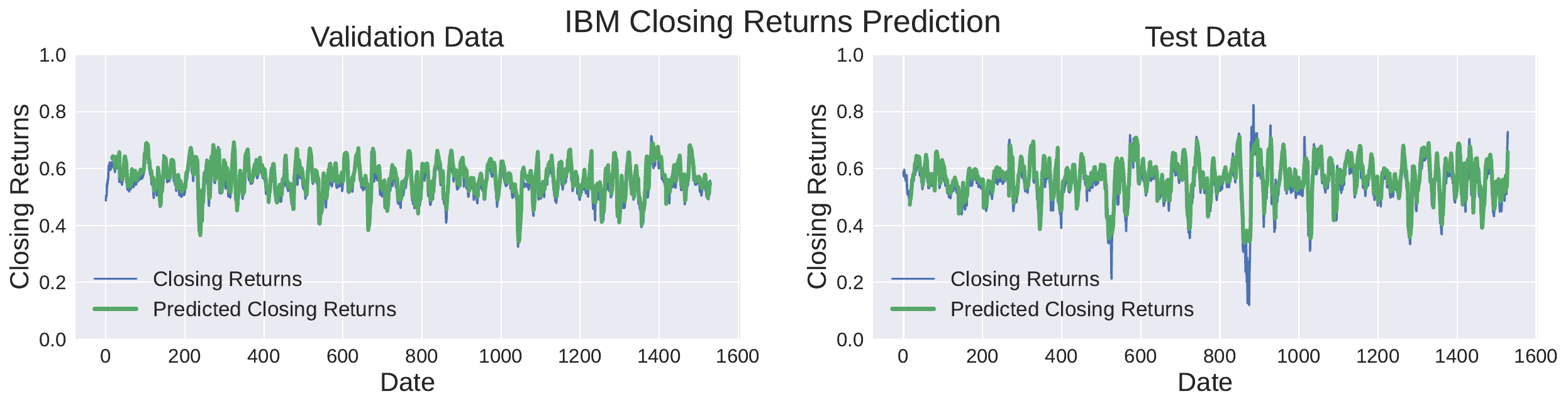}
         \caption{FedAvg}
         \label{fig:solo_IBM}
     \end{subfigure}
     \hfill
     \begin{subfigure}[b]{\columnwidth}
         \centering
         \includegraphics[width=0.92\columnwidth]{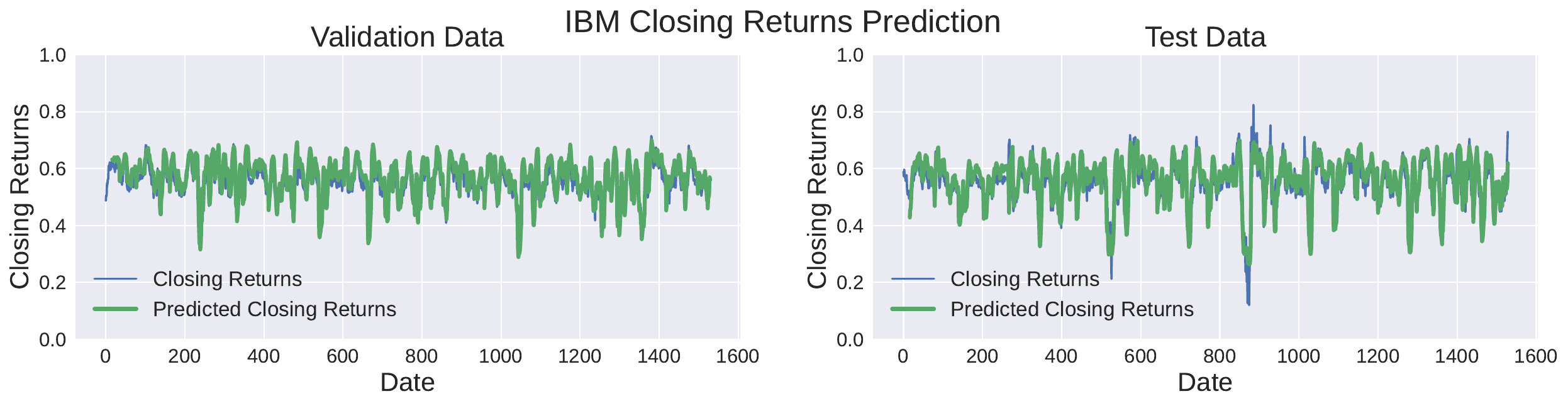}
         \caption{FedAtt (Proposed)}
         \label{fig:fedatt_IBM}
     \end{subfigure}
        \caption{Visualization of forecasting results on IBM dataset.}
        \label{fig:ibm_result}
\end{figure}

\subsection{Implementation details} We implement our models using the Tensorflow framework. We conduct the manual hyperparameter tuning for the best settings: batch size as 32, sequence length as 16, regarding a duration of 16 days, and for the transformer, we set the embedding dimension as 256 and employ 12 attention heads. We use the Adam optimizer with a default learning rate of 0.001. We conduct 10 training epochs for the decentralized SOLO method while conducting 10 global rounds with a single local epoch for both FedAtt and FedAvg. We conduct all the experiments on a single NVIDIA RTX 3080 GPU with 10GB memory. Followings are the evaluation metrics for the performance comparison with $M$ for the total sample size:
\begin{itemize}
    \item MSE: Mean Squared Error
    \begin{equation}
    MSE = \frac{1}{M}\sum_{i=1}^M(actual_i-forecast_i)^2
    \end{equation}
    
    \item MAE: Mean Absolute Error
    \begin{equation}
    MAE = \frac{1}{M}\sum_{i=1}^M|actual_i-forecast_i|
    \end{equation}
    
    \item MAPE: Mean Absolute Percentage Error
    \begin{equation}
    MAPE = \frac{1}{M}\sum_{i=1}^M|\frac{actual_i-forecast_i}{actual_i}|
    \end{equation}
\end{itemize}

\subsection{Experiment Results}

\begin{figure}[t]
     \centering
     \begin{subfigure}[b]{\columnwidth}
         \centering
         \includegraphics[width=0.92\columnwidth]{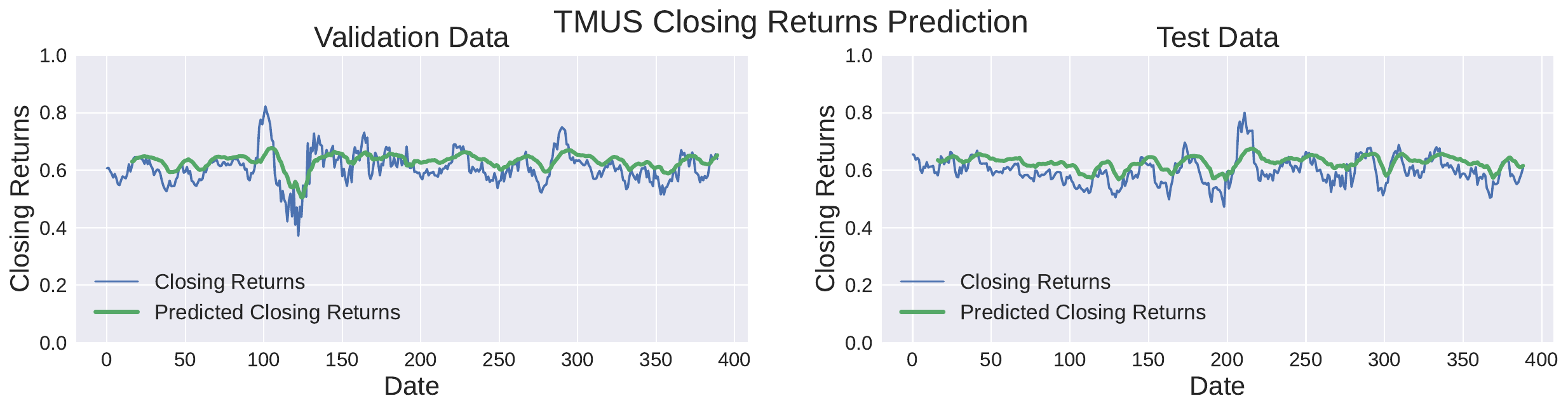}
         \caption{SOLO}
         \label{fig:solo_TMUS}
     \end{subfigure}
     \hfill
     \begin{subfigure}[b]{\columnwidth}
         \centering
         \includegraphics[width=0.92\columnwidth]{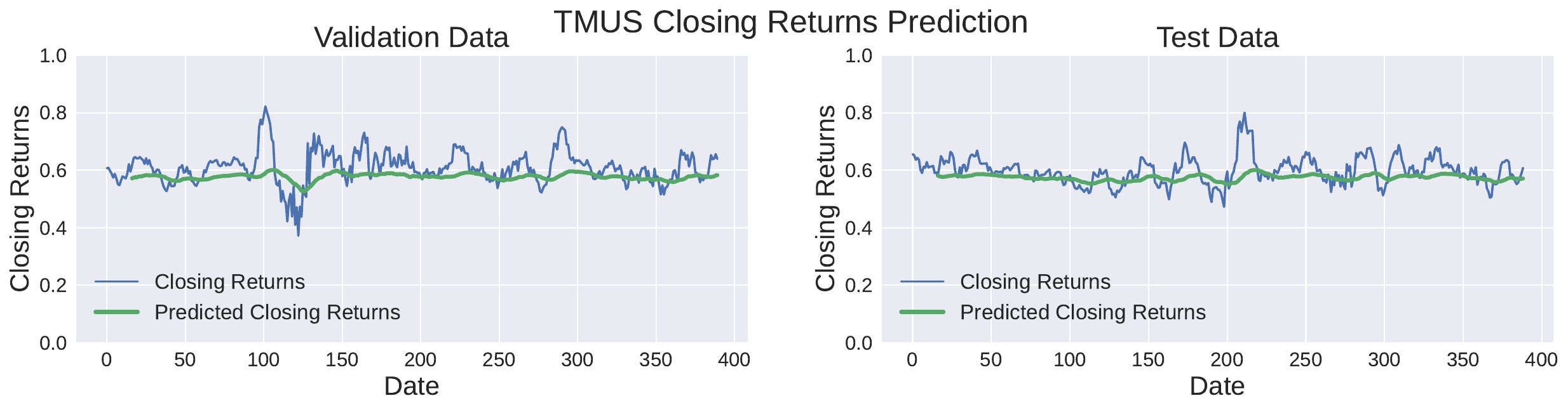}
         \caption{FedAvg}
         \label{fig:fedavg_TMUS}
     \end{subfigure}
     \hfill
     \begin{subfigure}[b]{\columnwidth}
         \centering
         \includegraphics[width=0.92\columnwidth]{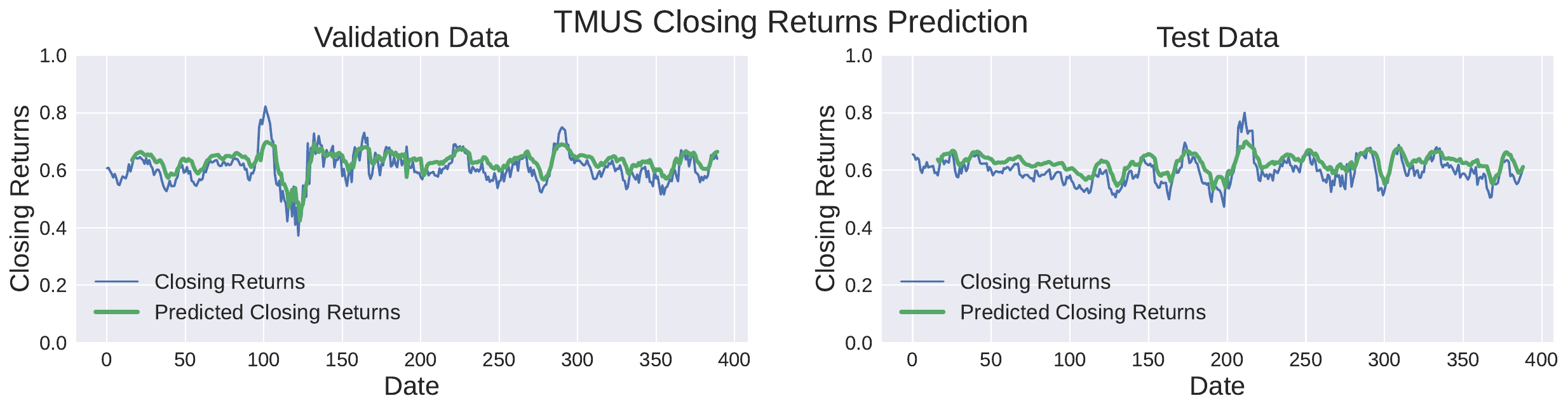}
         \caption{FedAtt (Proposed)}
         \label{fig:fedatt_TMUS}
     \end{subfigure}
        \caption{Visualization of forecasting results on TMUS dataset.}
        \label{fig:tmus_result}
\end{figure}

We conduct extensive experiments on the historical daily stock data retrieved from the Yahoo! Finance website and forecast the closing returns of each enterprise. Table.~\ref{metrics} shows the performance comparison on the test data for 5 of 45 enterprises. From the table, we can observe that our proposed scheme outperforms the other two baselines in most cases. We also provide the visualization of forecasting results on the validation and test sets for `International Business Machines Corporation (IBM)', and `T-Mobile US, Inc. (TMUS)' in Fig.~\ref{fig:ibm_result} and \ref{fig:tmus_result}, respectively.
In Fig.~\ref{fig:ibm_result}, the forecasting trends of all approaches are plotted perfectly with only a few differences. It is an expected result as the IBM dataset contains a large number of data points starting from the date `1962-01-02', which is a sufficient amount to train a high-performing model even solely with its private data.
Fig.~\ref{fig:tmus_result} indicates that our proposed FedAtt scheme outperforms the other baselines. It is because the TMUS dataset contains only a small number of data points starting from the date `2007-04-19'. Thus, when training a model locally in the decentralized setting, it cannot obtain peak performance as the number of data points is not enough to train a data-hungry transformer. Also, in the FedAvg scheme, the importance of each client is ignored, and thus, our proposed method outperforms the others with the help of an attentive aggregation mechanism.

\section{Conclusion} \label{conclusion}
In this work, we propose attentive federated transformers for time series stock forecasting, proving the fact that transformers can capture long-range dependencies and interactions among sequential data. We also explore the effectiveness of our proposed time series transformers in FL scenarios to enhance the forecasting accuracy while coping with data heterogeneity, scarcity, and privacy issues. Thus, we exploit the attentive federated learning (FedAtt) scheme to enable the collaborative training of our models, leveraging the distributed historical stock data of different enterprises. Empirical results on various stock market data show the superiority of our proposed scheme in comparison with the decentralized local training (SOLO) and the federated averaging (FedAvg) baselines. Thus, we can conclude from our findings that federated transformers are effective for time series forecasting tasks, considering our future direction on the data-intensive time series applications in medical and financial domains.

\bibliographystyle{IEEEtran}
\bibliography{main}

\end{document}